\documentclass[12pt,twocolumn]{article}
\usepackage{epsfig}
\usepackage{latexsym}
\usepackage{amssymb}
\usepackage{graphicx}
\usepackage{times}
\renewcommand{\baselinestretch}{0.9}

\makeatletter
\def\section{\@startsection{section}{1}{\z@}{3.5ex plus-.3ex minus-.2ex}%
{1.5ex plus.3ex}{\small\bfseries}}
\renewcommand{\@makecaption}[2]{\footnotesize Fig.\hspace{-.2em}#1.\ \ #2}
\makeatother

\begin{document}
\twocolumn[\title{%{\normalsize Version Date: \today}\\%\vspace{.5cm}
\vspace*{-2.5cm}\Large Influence of Correlated Disorder Potentials on 
the Levitation of Current Carrying States in the Quantum Hall Effect}
\author{\vspace{.2cm}{\large Th. Koschny and L. Schweitzer\footnotemark[1]}  
\\[1.7\baselineskip]
{\footnotesize\itshape Physikalisch-Technische Bundesanstalt, Bundesallee 100,
38116 Braunschweig, Germany}}
\date{}
\maketitle
\vspace{-.5cm}
\rule{\textwidth}{.5pt}

\vspace{-.5cm}
\renewcommand{\baselinestretch}{0.9}
\small
\begin{abstract}
The disorder driven quantum Hall to insulator transition is
investigated for a two-dimensional lattice system. 
We consider a Gaussian correlated random potential, study 
the behaviour of the current carrying states and trace their 
energetical position when the disorder strength is increased.
Our results qualitatively resemble those obtained previously
for exponentially correlated disorder potentials.
We find both the downward movement of the anti-Chern states as 
well as the floating up of the Chern states across the Landau gap
which is sometimes masked by the global broadening of the tight binding band.

\end{abstract}

\vspace{.2cm}

\rule{\textwidth}{.5pt}
\vspace{.35cm}
]
\footnotetext[1]{Corresponding author. Fax: +49\,531\,5928106; 
email: Ludwig.Schweitzer@ptb.de.}

\makeatletter
\def\section{\@startsection{section}{1}{\z@}{3.5ex plus-.3ex minus-.2ex}%
{1.5ex plus.3ex}{\normalsize\bfseries}}
\makeatother
\renewcommand{\baselinestretch}{1.}
\normalsize
\section{Introduction}
The disorder driven quantum Hall to insulator transition 
once again became a topic of active research recently. 
The theoretical activities undertaken over the past years resulted in
a controversial debate concerning the fate of the current carrying 
states \cite{YB96,LXN96,XLSN96,SW97,SW98,PBS98,YB99}. 
Likewise, the experimental results reported all along remain 
inconclusive \cite{SKD93,JJWH93,Wea94,KMFP95,GJJ95,Sea97,LCSL98,Hea00}.
In numerical studies using uncorrelated random disorder potentials, 
it was found that in contrast to the continuum model, where the 
critical states are known to float up in energy \cite{Khm84,Lau84,And84},
the current carrying states are annihilated by anti-Chern states
originating from the band centre in the lattice model \cite{LXN96,SW97,PBS98}.
It has also been claimed \cite{SW97,SW98} that such a scenario should be 
able to explain the direct QH to insulator transitions from higher Hall
plateaus with filling factor $\nu > 2$, which apparently have been 
observed in several experiments. 
%Also, a new universality class has
%been attributed to these transitions \cite{Sea97,SW98}.
However, according to the proposed global phase diagram \cite{KLZ92}, 
which is based on the levitation picture, such transitions should not 
be allowed. 

In a recent paper we have shown \cite{KPS01} that 
the lattice model can exhibit similar properties as the continuum
model if long range correlated random disorder potentials are considered. 
Using exponentially correlated random potentials to model the 
intrinsic scattering potentials and other imperfections that may
influence the movement of the charge carriers in real 2DEGs, we
found that the floating up in energy can already be observed for correlation
lengths larger than about half the lattice constant. In particular,
with increasing disorder the floating up of the critical energy  
across the Landau gap without merging has been demonstrated.
The latter is in contrast to the results of a numerical study 
\cite{SWW00} where Gaussian correlated disorder potentials were used. 
In the present paper we investigate whether the observed discrepancy 
is due to the differences in the correlated disorder potentials considered.
Therefore, in the following we also use Gaussian correlated disorder 
potentials and calculate numerically the disorder dependence of the 
density of states and investigate the finite size scaling of the 
localisation length from which the energetical position of the 
current carrying states can be extracted.

\section{Model and Method}
The QHE system is described by a disordered two-dimensional single 
band lattice model \cite{SKM84,KPS01}, 
$H=\sum_{m}w_m^{}c_{m}^{\dagger}c_m^{}+
\sum_{<mn>} (V_{nm}c_{m}^{\dagger}c_n^{}+
V_{mn}c_{n}^{\dagger}c_m^{})$, 
where $c_{m}^{}, c_m^{\dagger}$ are the creation and annihilation
operators of a particle at site $m$, respectively. 
The perpendicular magnetic field is incorporated
via the complex phases of the transfer terms, $V_{mn}=V \exp(ib_{mn})$, 
between neighbouring sites $m,n$ on the lattice.
In the Landau gauge,
$b_{mn} = \pm 2\pi (p/q) (\vec{r}_m\cdot\vec{e}_y)/a$, if
$\vec{r}_n=\vec{r}_m\pm a\vec{e}_x$, and $b_{mn} = 0$ else, 
where $\vec{r}_m$ is the position of site $m$ and $\vec{e}_x$, 
$\vec{e}_y$ are unit vectors pointing in the $x$ and $y$ directions.
We choose $p/q=1/8$ flux quanta $h/e$ per plaquette $a\times a$, where
$a$ is the lattice constant.
The Gaussian correlated disorder potentials $w_m$ are generated 
from uncorrelated random variables $\varepsilon_n$ associated with each 
lattice site $n$ by local
averaging using a Gaussian weighting function, 
$w_m=1/N\ \sum_n \varepsilon_n\,\exp(-|m-n|^2/\eta^2)$, with  
correlation length $\eta$ and some normalisation factor $N$.
This leads to a spatial correlation function decaying qualitatively 
like $\langle w_m w_n\rangle \sim\exp(-|m-n|^2/2\eta^2)$.
The probability density of the $\varepsilon_n$ is taken to be 
$P(\varepsilon_n)=1/(2W)$ in the range $-W\le\varepsilon_n\le W$,
and zero else.

\begin{figure}
\centering
\includegraphics[width=7.7cm]{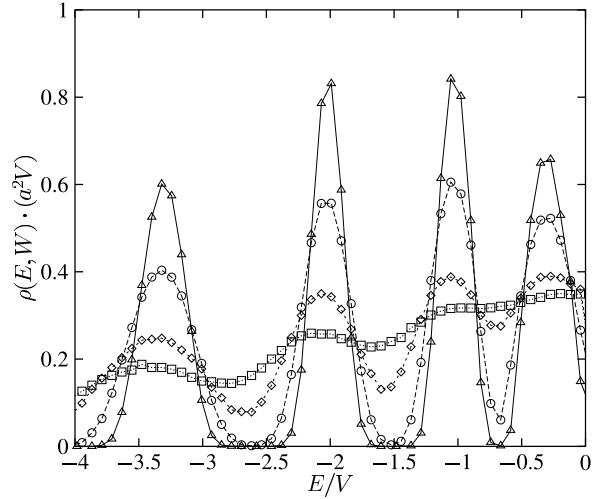}
\smallskip
\caption{The density of states $\rho(E,W)$ versus energy for 
disorder strengths $W/V=1.0, 1.5, 2.5, \textnormal{and\ }
3.5$. The system size is $M/a=48$ and the lines simply connect the 
calculated data points.}
\label{dos1}
\end{figure}

The density of states $\rho(E,W)$ is calculated from the distribution of
energy eigenvalues which are obtained by direct diagonalisation of square 
systems averaged over 60 realisations of the correlated disorder potentials.
The exponential localisation length 
$\lambda_M^{-1}(E,W)=-\lim_{L\to\infty} (2L)^{-1}
\ln\textnormal{Tr}\{|G_{1L}|^2\}$ 
is calculated numerically by means of a recursive Green function method 
\cite{MK83,SKM84}. 
$G\equiv(E-H+i\gamma)^{-1}$ is the total Green function and $G_{1L}$
is defined on the subspace of the first and last columns on the lattice.
The system's length and width are $L$ and $M$, respectively.
Both, the density of states and the localisation length were computed
for various system sizes, disorder strengths, and correlation lengths 
of the disorder potentials.  
In what follows, we restrict the presentation of our results to those
data obtained for correlation parameter $\eta/a=1.0$. 
The same value has been utilised in Ref.~\cite{SWW00}.

\section{Results and Discussion}
The movement of the energetic position of the current carrying states
induced by a change of the disorder strength is obscured by the global
broadening of the tight binding band with increasing disorder.
To distinguish this effect, which in turn shifts the Landau bands 
outwards, from the true floating of the critical states, 
the density of states $\rho(E)$ is calculated first for
several disorder strength in the range $0.5\le W/V \le 5.0$.
As an example, the result of $\rho(E)$ for disorder strength 
$W/V=1.0, 1.5, 2.5, \textnormal{and\ } 3.5$ is plotted in Fig.~\ref{dos1}. 
Due to the symmetry with respect to $E/V=0$, 
only the lower half of the total band is shown. 
With increasing disorder the sub-bands broaden and shift slightly
outwards. The effect is stronger near the band edge. Similar curves
were obtained for further disorder strengths.
The position of the sub-band peaks were extracted from a 
curve-fitting procedure using Gaussian or power-like shapes for the 
particular Landau bands, exploiting the normalisation condition for each 
single band and tracing the fit parameters from lower to higher disorder. 

\begin{figure}
\centering
\includegraphics[width=7.7cm]{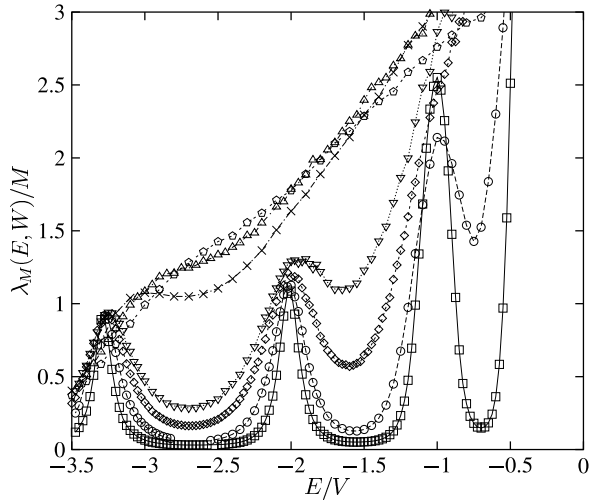}
\smallskip
\caption{The energy dependence of the normalised localisation length 
$\lambda_M(E,W)/M$ for disorder strengths $W/V=1.5$, $2.5$, $3.5$, 
$4.0$, $5.5$, $6.0$, and $6.5$. 
The system length is at least $L/a=3\cdot 10^4$ and the width $M/a=48$.}
\label{lambdaE}
\end{figure}

The energy dependence of the normalised localisation length 
$\lambda_M(E,W)/M$ has been calculated for disorder strengths in the range 
$1.0\le W/V\le 6.5$.  Curves for $W/V=1.5$, $2.5$, $3.5$, $4.0$, $5.5$,
$6.0$, and $6.5$  are plotted in Fig.~\ref{lambdaE}, whereby 
the data points shown are averages over 30 disorder realizations. The
system width is $M/a=48$ and the length varies between $L/a=3\cdot 10^4$
and $L/a=5\cdot 10^5$ depending on the varying convergence for 
different energy and disorder strength. For fixed system width $M$, 
$\lambda_M(E,W)/M$ increases with increasing disorder strength. 
Associated with this increase is a shift of the peak position, 
which corresponds to the floating up of the current carrying state 
to higher energies.    

\begin{figure}
\centering
\includegraphics[width=7.7cm]{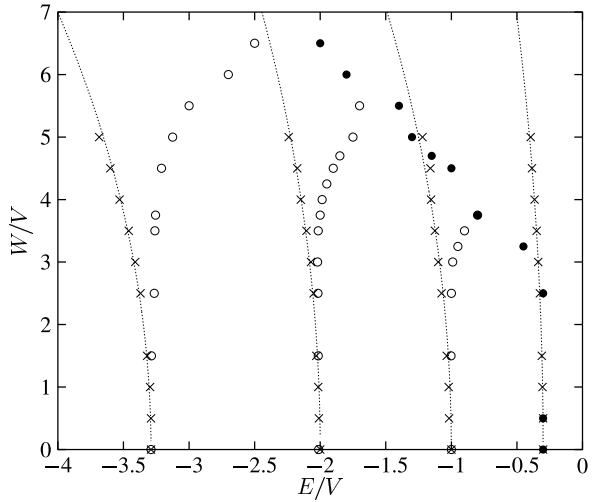}
\smallskip
\caption{The position of the sub-band peaks in the density of states 
$\rho(E,W)$ ({\footnotesize $\times$}) and the position of the normalised
localisation length $\lambda_M(E,W)/M$ ({\large $\circ$}) 
within the disorder-energy plain. The anti-Chern states are depicted 
by ({\large $\bullet$}). The dotted line is an empirical fit 
$W_p(E)=A  (E_n-E)^{1/2}$ to the peak position of the sub-bands.}
\label{WE_dosl}
\end{figure}

In Fig.~\ref{WE_dosl} the energy and disorder dependence of
$\rho(E,W)$ and $\lambda_M(E,W)/M$ are combined. The position of 
the sub-band peaks of the density of states ({\footnotesize $\times$}) 
is shown within the disorder-energy plane. With increasing disorder, 
the density of state peaks shift to lower energy. This movement 
is well fitted by the empirical relation $W_p(E)=A  (E_n-E)^{1/2}$
which corresponds to the observation that the bandwidth of the total
band increases with disorder $\sim W^2$. The $E_n$ denote the energies
of the sub-bands (Landau levels) for $W/V=0$.
In addition to the sub-band peaks, the position of the current 
carrying states ({\large $\circ$}) are plotted also in Fig.~\ref{WE_dosl}. 
The energies of the critical states were extracted from a finite size
scaling analysis combined with a similar curve-fitting procedure as
used for the total density of states above.
For each disorder strength $\lambda_M(E,W)/M$ decreases for localised
states with increasing system size while it stays constant in the 
thermodynamic limit only at those
energies that correspond to the current carrying states.
  
For small disorder $W$ both the energy of the peaks in the density of
states and the energy of the critical states coincide. 
However, with increasing disorder strength, the curves move apart. 
The floating of the energy of the critical states in the opposite
direction as the sub-bands, almost across the lowest Landau gaps, 
is clearly seen. The downward movement of the anti-Chern state 
({\large $\bullet$}) can easily
be followed until it is finally annihilated by the Chern state of the
lowest sub-band at a disorder strength of $W/V\sim 7$. 
This also defines the cross-over from the quantum Hall liquid state to
the insulator.

Comparing our results with those published previously for exponentially
correlated disorder potentials \cite{PS01,KPS01}, we notice that the
behaviour of the current carrying states is qualitatively similar. 
With increasing
correlations the position where the anti-Chern and the last Chern
state meet is shifted to higher energies. Therefore, it is to be
expected that for stronger correlations only the floating
up of the energy of the current carrying states is observed as in
the continuum model.
The apparent merging and subsequent joint floating of the current
carrying states as reported \cite{SWW00} for similar Gaussian correlated 
disorder potentials was not observed in our investigation. However,
this may look different for smaller magnetic field as applied in 
\cite{SWW00} because the distinction of clustered critical states 
becomes increasingly problematic in finite size studies.

In conclusion, the disorder driven quantum Hall to insulator transition 
in a lattice model with Gaussian correlated random potentials has been
investigated. 
For a correlation length $\eta/a=1$ we have shown that 
the energy of the lowest current carrying state floats up in energy 
across the Landau gap when the disorder strength is increased. 
This is similar to the levitation scenario proposed for the continuum 
model. In view of these results, the proposed explanation \cite{SWW00}
for the apparent direct transitions to the insulator 
from plateaus with $\nu>2$ as being due to the downward movement of
the anti-Chern states is unlikely. We believe that the direct 
transitions reported in experiments are either unresolved single one by one
transitions or have to be accounted for by electron-electron interactions.

%\bibliographystyle{permyear_prsty}
%\bibliographystyle{elsart-num}
%\bibliography{ludwig,papers2000_database,qhe}

\begin{thebibliography}{10}\setlength{\itemsep}{-3.0pt}
\footnotesize
\bibitem{YB96}
K. Yang and R.~N. Bhatt, Phys. Rev. Lett. {\bf 76},   (1996) 1316.

\bibitem{LXN96}
D. Liu, X. Xie, and Q. Niu, Phys. Rev. Lett. {\bf 76},   (1996) 975.

\bibitem{XLSN96}
X.~C. Xie, D.~Z. Liu, B. Sundaram, and Q. Niu, Phys. Rev. B {\bf 54},   (1996)
  4966.

\bibitem{SW97}
D.~N. Sheng and Z.~Y. Weng, Phys. Rev. Lett. {\bf 78},   (1997) 318.

\bibitem{SW98}
D.~N. Sheng and Z.~Y. Weng, Phys. Rev. Lett. {\bf 80},   (1998) 580.

\bibitem{PBS98}
H. Potempa, A. B\"aker, and L. Schweitzer, Physica B {\bf 256-258},   (1998)
  591.

\bibitem{YB99}
K. Yang and R.~N. Bhatt, Phys. Rev. B {\bf 59},   (1999) 8144.

\bibitem{SKD93}
A.~A. Shashkin, G.~V. Kravchenko, and V.~T. Dolgopolov, JETP Lett. {\bf 58},
  (1993) 220.

\bibitem{JJWH93}
H.~W. Jiang, C.~E. Johnson, K.~L. Wang, and S.~T. Hannahs, Phys. Rev. Lett.
  {\bf 71},   (1993) 1439.

\bibitem{Wea94}
T.~K. Wang {\it et~al.}, Phys. Rev. Lett. {\bf 72},   (1994) 709.

\bibitem{KMFP95}
S.~V. Kravchenko, W. Mason, J.~E. Furneaux, and V.~M. Pudalov, Phys. Rev. Lett.
  {\bf 75},   (1995) 910.

\bibitem{GJJ95}
I. Glozman, C.~E. Johnson, and H.~W. Jiang, Phys. Rev. Lett. {\bf 74},   (1995)
  594.

\bibitem{Sea97}
D. Shahar {\it et~al.}, Phys. Rev. Lett. {\bf 79},   (1997) 479.

\bibitem{LCSL98}
C.~H. Lee, Y.~H. Chang, Y.~W. Suen, and H.~H. Lin, Phys. Rev. B {\bf 58},
  (1998) 10629.

\bibitem{Hea00}
M. Hilke {\it et~al.}, Phys. Rev. B {\bf 62},   (2000) 6940.

\bibitem{Khm84}
D.~E. Khmelnitskii, Phys. Lett. {\bf 106A},   (1984) 182.

\bibitem{Lau84}
R.~B. Laughlin, Phys. Rev. Lett {\bf 52},   (1984) 2304.

\bibitem{And84}
T. Ando, Journal of the Physical Society of Japan {\bf 53},   (1984) 3126.

\bibitem{KLZ92}
S. Kivelson, D.-H. Lee, and S.-C. Zhang, Phys. Rev. B {\bf 46},   (1992) 2223.

\bibitem{KPS01}
T. Koschny, H. Potempa, and L. Schweitzer, Phys.\ Rev.\ Lett. {\bf 86},
  (2001) 3863.

\bibitem{SWW00}
D.~N. Sheng, Z.~Y. Weng, and X.~G. Wen, Cond-mat/0003117  (2000) .

\bibitem{SKM84}
L. Schweitzer, B. Kramer, and A. MacKinnon, J. Phys.\ C {\bf 17},   (1984)
  4111.

\bibitem{MK83}
A. MacKinnon and B. Kramer, Z. Phys. B {\bf 53},   (1983) 1.

\bibitem{PS01}
H. Potempa and L. Schweitzer, Physica B {\bf 298},   (2001) 52.

\end{thebibliography}

\end{document}